\documentclass[twocolumn]{article}
\usepackage[margin=0.6in]{geometry}

\usepackage{amsmath}

\usepackage{filecontents}
\usepackage{booktabs}
\usepackage{caption}
\usepackage{subcaption}
\usepackage{tablefootnote}

\usepackage{cite}

\usepackage{amsthm}
\theoremstyle{definition}

\usepackage{xspace}

\newcommand{\work}[1]{{\small \sf #1}}

\newcommand{\sysname}{{\small \sf STRIDE}\xspace}
\newcommand{\dirty}{{\small \sf DIRTY}\xspace}
\newcommand{\varbert}{{\small \sf VarBERT}\xspace}

\newcommand{\smallsec}[1]{
{\small
\medskip
\noindent\textbf{\textsf{#1}}
}
}

\usepackage{tabularx}
\usepackage{multirow}
\usepackage{graphicx}

\usepackage{svg}
\usepackage{float}

\newcommand{\ie}{i.e.,\xspace}

\definecolor{tblue}{RGB}{0, 90, 181}
\definecolor{tred}{RGB}{220, 50, 32}
\definecolor{dkred}{RGB}{150, 0, 0}

\usepackage{hyperref}

\usepackage{authblk}

\begin{document}

\date{}

\title{STRIDE: Simple Type Recognition In Decompiled Executables}

\author[1]{Harrison Green}
\author[2]{Edward J. Schwartz}
\author[1]{Claire Le Goues}
\author[1]{Bogdan Vasilescu}

\affil[1]{Carnegie Mellon University. \texttt{\{harrisog, clegoues, vasilescu\}@cmu.edu}}
\affil[2]{Carnegie Mellon University Software Engineering Institute. \texttt{eschwartz@cert.org}}

\maketitle

\begin{abstract}

Decompilers are widely used by security researchers and developers to reverse engineer executable code. While modern decompilers are adept at recovering instructions, control flow, and function boundaries, some useful information from the original source code, such as variable types and names, is lost during the compilation process. Our work aims to \textit{predict} these variable types and names from the remaining information.

We propose \sysname, a lightweight technique that predicts variable names and types by matching sequences of decompiler tokens to those found in training data. We evaluate it on three benchmark datasets and find that \sysname achieves comparable performance to state-of-the-art machine learning models for both variable retyping and renaming while being much simpler and faster. We perform a detailed comparison with two recent SOTA transformer-based models in order to understand the specific factors that make our technique effective. We implemented \sysname in fewer than 1000 lines of Python and have open-sourced it under a permissive license at \url{https://github.com/hgarrereyn/STRIDE}.

\end{abstract}

\section{Introduction}
\label{sec:introduction}

To understand the intricate workings of compiled executables, reverse engineers turn to \textit{decompilers}: systems that can recover and/or regenerate the patterns found in high-level source code. While programming languages are designed to be written and read by humans, compiled executables bear no such constraints---designed instead to be interpreted and executed by computers. The  \textit{compilation} process therefore loses information whereby high level constructs present in source code are mangled and often elided during the translation into an executable format. Indeed, it is the job of the decompiler to undo this information loss, identifying function boundaries, reconstructing high-level control flow, and (the subject of this paper) recovering the \textit{types} and \textit{names} of variables.

\begin{figure}
    \resizebox{\columnwidth}{!}{
        \fbox{
            \includegraphics[trim=0 8 0 8]{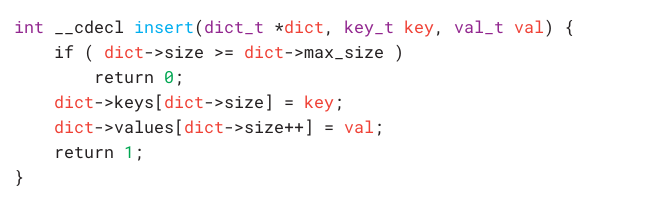}
        }
    }
    \caption{Hex-Rays decompilation (with names and types)}
    \label{fig:hexrays-types}
\end{figure}

\begin{figure}
    \resizebox{\columnwidth}{!}{
        \fbox{
            \includegraphics[trim=0 8 0 8]{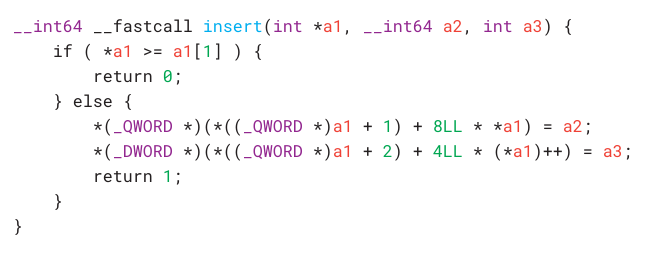}
        }
    }
    \caption{Hex-Rays decompilation (without names and types)}
    \label{fig:hexrays-stripped}
\end{figure}

While modern decompilers are powerful, accurately naming and typing variables remains challenging. As an example, consider Figures \ref{fig:hexrays-types} and \ref{fig:hexrays-stripped}.
These figures show the same function decompiled by Hex-Rays~\cite{hexrays},
\footnote{\href{https://www.hex-rays.com/products/decompiler/}{https://www.hex-rays.com/products/decompiler/}} 
a popular decompiler used in industry. \autoref{fig:hexrays-stripped} shows the default decompilation and \autoref{fig:hexrays-types} shows the decompilation when Hex-Rays has been provided with the correct variable types and names via DWARF debug metadata. Using this metadata, Hex-Rays is able to correctly set types and names and produce output that closely resembles the original source code (not shown). Without this information however, Hex-Rays resorts to displaying raw pointer arithmetic with unhelpful names like \texttt{a1} and \texttt{a2}.

Unfortunately, \autoref{fig:hexrays-stripped} is the reality for most real-world applications. Typically reverse engineers \textit{only} have access to stripped executables and not source code or debug information, such as when: examining malware samples~\cite{yakdan:2016,votipka:2020,durfina:2013,vanemmerik:2007,yakdan2015no,burk2022decomperson}, looking for security vulnerabilities in closed-source software~\cite{mantovani2022convergence,vanemmerik:2007,votipka:2020,kalle2019clik,burk2022decomperson,liu2020far,yakdan2015no}, reverse engineering legacy code~\cite{liu2020far,emmerik2004using,jaffe:2018,cifuentes1996partial,fokin:2011:smartdec,mycroft:1999}, or competing in cybersecurity competitions~\cite{burk2022decomperson,chapman2014picoctf,liu2020far,burns2017analysis,song:2015}. Techniques which can \textit{predict} this missing information and restore correct variable names and types would therefore improve the effectiveness of decompilers across a wide array of tasks, as well as the experience of the reverse engineers using them.

\textit{Variable type inference}, the task of predicting \textit{types} for variables, is a long-studied problem in the context of decompilation \cite{typeinference_survey}. Prior work runs the gamut from principled static-analysis~\cite{lee2011tie, yakdan2015no,schwartz:2013,secondwrite,retypd} to dynamic techniques~\cite{lin2010automatic,howard}, handwritten rules~\cite{hexrays,osprey}, and most recently, machine learning techniques such as conditional random fields~\cite{debin}, neural models trained on embeddings~\cite{typeminer,cati}, and transformers trained on decompiler tokens~\cite{chen2022augmenting}.

\textit{Variable renaming}, the related task of assigning meaningful \textit{names} to variables has also been explored in recent years. Most prior work has adapted machine learning models derived from the field of natural language processing, including conditional random fields~\cite{debin}, gated graph neural networks (GGNNs) and long short-term memory networks (LSTMs)~\cite{lacomis2019dire}, and transformers~\cite{direct,chen2022augmenting,hext5,varbert}.

Reflecting on recent work in software engineering, there is a clear trend towards increasingly more sophisticated neural architectures and / or increasingly larger (language) models, including for decompilation-related applications~\cite{hudegpt, xu2023lmpa, tan2024llm4decompile}. And while we are excited about all these advances, we are also cautious---are ever-larger models really the solution to all problems, and are we really using that expensive-to-train and expensive-to-operate infrastructure most effectively?
In this paper, we demonstrate that perhaps the answer is `no' by
showing that a much simpler non-neural system, that mimics how a human reverse-engineer might approach the variable type inference and variable renaming problems, can achieve comparable performance to state-of-the-art transformer-based models, while being both faster and cheaper to train.

The main intuition behind our approach is that the most relevant information for predicting the name or type of a variable from incomplete decompilation is located immediately next to places where this variable is mentioned. In other words, to predict the type of \texttt{a1}, it is sufficient to look near where \texttt{a1} appears in the decompilation. These are the locations where the variable is \textit{used}, and therefore the most relevant places to find \textit{contextual hints} about its name or type.

We generalize this concept of \textit{contextual hints} with N-grams, matching the sequence of tokens before and after a variable to sequences found in our training data. In other words, if we find a variable that appears in the same context as a variable in our training data---i.e., a sequence of decompiler tokens matches exactly---we gain confidence that these variable have the same type and/or name. If we are able to match \textit{longer sequences} around this variable or match \textit{more instances} of the variable within the same function, we gain more confidence still that the match is correct.

We implemented these ideas in our system \sysname ({\sf \underline{S}imple \underline{T}ype \underline{R}ecognition \underline{I}n \underline{D}ecompiled \underline{E}xecutables}), available open source. \footnote{\href{https://github.com/hgarrereyn/STRIDE}{https://github.com/hgarrereyn/STRIDE}}
Extensive evaluation shows that \sysname is able to match (and often surpass) the performance of recent SOTA systems at the tasks of variable retyping and renaming. In particular, \sysname achieves 66.4\% accuracy on variable retyping (14.1\% improvement) and 56.2\% accuracy on variable renaming (4.9\% improvement) on the not-in-train split of the \work{DIRT} benchmark dataset used by prior work. Furthermore, compared to \varbert, a SOTA model for variable renaming, \sysname achieves an average of 1.5\% higher accuracy at variable renaming on the \work{VarCorpus} dataset across two decompilers and four optimization levels---and it does so while making predictions 5 times faster, without using a GPU, and without pre-training on source code.

In summary, our contributions are the following:
\begin{itemize}
  \item We propose \sysname, a simple approach to variable renaming and retyping using N-grams that acts as a post-processor to decompilers. Our approach is motivated by our hypothesis that the most relevant context for a variable's name and type consists of the surrounding context where it is referenced.
  \item We compare the performance of \sysname at variable \textit{retyping} and \textit{renaming} to the state-of-the-art models \dirty and \varbert on three benchmark datasets. %
  \item We perform a series of evaluations comparing \sysname to \dirty and \varbert to understand how architectural differences affect task performance and discuss the potential for applying these systems in a real-world setting.
\end{itemize}

\section{Related Work}

\subsection{Predicting Types}

\textit{Variable type inference}, the task of assigning \textit{types} to variables, is a well studied problem in the context of decompilation, however prior work is varied in both methodology and objective \cite{typeinference_survey}. Many foundational works in this area are concerned with purely the \textit{structure} and \textit{shape} of the types, disambiguating between a handful of primitive types and arrays, without attempting to identify user-defined structs or name specific fields.

Systems such as \work{TIE}~\cite{lee2011tie, yakdan2015no,schwartz:2013}, \work{SecondWrite}~\cite{secondwrite}, and \work{Retypd}~\cite{retypd} take a principled static-analysis approach, lifting the machine code into an intermediate language to reason over its semantics. Alternatively, \work{REWARDS}~\cite{lin2010automatic} uses a dynamic technique, observing when variables are passed to known \textit{sinks} such as syscalls and retroactively assigning the types based on the signature. \work{HOWARD}~\cite{howard} improves upon this work by observing dynamic \textit{access patterns} in traces to identify structures like nested arrays.

Recently, several works have applied probabilistic techniques to this area. \work{DEBIN}~\cite{debin} uses Conditional Random Fields to classify variables among 17 primitive types. \work{TypeMiner}~\cite{typeminer} and \work{CATI}~\cite{cati} extract local features near the usage sites of variables and train machine learning models to classify them. \work{OSPREY}~\cite{osprey} uses a probabilistic technique with a set of handwritten rules to assign types to variables.

While these systems take quite varied approaches, they are generally designed to be conservative, predicting one of a handful of primitive types only when there is clear evidence. For example these systems may focus on the disambiguation of \texttt{float} vs. \texttt{int} vs. \texttt{char *}. This level of type prediction is useful in the early stages of decompilation in order to decide how to lay out code---and it is important to be conservative. However, with these techniques, reasoning about more complex types, for example large structs, is much harder. Indeed, human reverse engineers often use intuition about what types \textit{should} look like in order to extrapolate and fill in gaps when information is missing.

Recently, the authors of \work{DIRTY}~\cite{chen2022augmenting} demonstrated that using a statistical method of type classification---drawing from a large library of learned types rather than a fixed set---can be an effective strategy. Rather than cobbling a struct together piece by piece, \dirty uses a transformer-based machine learning model to identify instances of known types based on the surrounding context. This approach brings two large benefits: 1. \dirty can extrapolate probabilistically, recognizing types even if only a portion of the structure is used within a given function. 2. \dirty predicts full \textit{semantic types} with named fields rather than simply \textit{structural types}, \ie it can predict \texttt{struct Point {int x; int y}} instead of the anonymous type \texttt{struct {int; int}}. 

Our work continues in this lineage of statistical type recognition, but shows that much simpler (conceptually and to train) models are not only sufficient, but more accurate.

\subsection{Predicting Names}

The task of \textit{variable renaming} is similarly well-studied. Unlike \textit{variable retyping}, variable names cannot be deduced through pure static analysis---indeed, names have no direct bearing on the execution of the program. Thus, most recent works treat this problem as either as a generative modeling task or a large classification problem, trying either to generate or identify the \textit{most likely} name for a particular variable.

\work{DEBIN}~\cite{debin} (also mentioned previously) uses Conditional Random Fields to predict variable names. The most similar prior work in this area however are the systems \work{DIRE}~\cite{lacomis2019dire}, \work{DIRECT}~\cite{direct}, \work{DIRTY}~\cite{chen2022augmenting}, \work{HexT5}~\cite{hext5}, and \work{VarBERT}~\cite{varbert} which apply neural language models to the output of decompilation. We briefly summarize their contributions in the rest of this section.

\work{DIRE}~\cite{lacomis2019dire} combines lexical and structural information from the decompiler to make predictions. Specifically, a bi-directional long short-term memory (LSTM) network is first applied to the raw decompiler tokens to generate a \textit{lexical embedding}. To capture structural information, the technique applies a gated graph neural network (GGNN) to the abstract syntax tree produced by the decompiler to generate a \textit{structural embedding}. These embeddings are then processed by a third network, the \textit{decoder}, which is also an LSTM. 

\work{DIRECT}~\cite{direct} applies a BERT model to the task of variable renaming. Decompiler output is split into 512-token chunks and passed through a BERT decoder model to fixup variable names.

\dirty~\cite{chen2022augmenting} fuses the problems of variable name and type prediction with a multi-task transformer-based decoder network. The authors feed the raw decompiler output to a lexical encoder network and then apply these embeddings to the decoder network, predicting variable names and types in interleaved steps. Type predictions have a soft mask applied based on a \textit{data layout encoding} that considers factors like predicted variable size and location.

\work{HexT5}~\cite{hext5} fine-tunes the CodeT5 model, a transformer-based architecture trained for source code understanding and generation, on several downstream decompilation-related tasks including variable renaming.

\varbert~\cite{varbert} is the newest and current state-of-the-art approach for decompiler variable renaming. The authors apply a BERT model to variable renaming by first pre-training on a large corpus of source code through masked language modeling (MLM) and constrained masked language modeling (CMLM). They then fine-tune each model on decompilation, again with constrained masked language modeling.

While our work continues in this tradition of using decompiler tokens as input, we apply much simpler techniques designed to mirror human intuition on the same tasks.

\section{Reverse Engineering with Decompilation}
\label{sec:rev}

Next we provide background and motivation for our proposed system \sysname, giving an example of how human reverse engineers might use the information present in (incorrect) decompilation to arrive at new variable types and names.

\smallsec{Decompilation.}
Decompilers take compiled executables and generate \textit{decompilation}. This decompilation resembles source code but does not necessarily conform to language rules or recompile---indeed its primary purpose is to inform the reverse engineer. While decompilation without accurate types may be confusing and verbose, as in \autoref{fig:hexrays-stripped}, it still provides a lot of useful hints. In fact, \dirty~\cite{chen2022augmenting} demonstrated that a transformer-based model trained \textit{only} on these decompiler tokens could achieve SOTA performance (at that time) at variable renaming and retyping. Experienced human reverse engineers are capable of manually recovering variable types and names by looking at this decompilation, so why can't automated techniques as well?

\begin{figure}
    \resizebox{\columnwidth}{!}{
        \includegraphics[clip=true,trim=12 14 12 12]{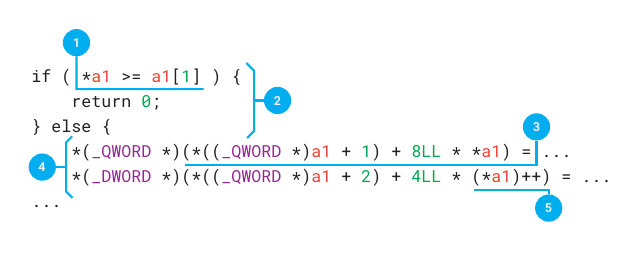}
    }
    \caption{Decompilation provides contextual hints even with wrong names and types.}
    \label{fig:hints}
\end{figure}

\begin{figure}
    \resizebox{\columnwidth}{!}{
        \includegraphics[clip=true,trim=12 10 18 5]{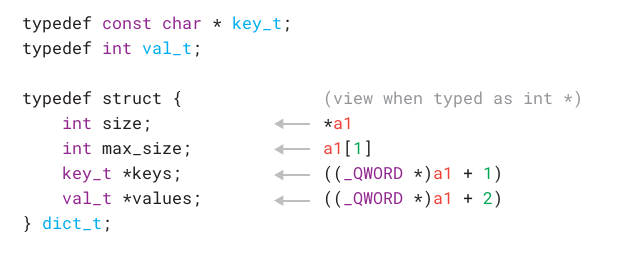}
    }
    \caption{Running example ground truth types. When a \texttt{dict\_t *} is mistyped as an \texttt{int *}, the field accesses look different.}
    \label{fig:types}
\end{figure}

\smallsec{Variable Context Provides Hints.}
So what exactly can we learn from this incomplete decompilation? In \autoref{fig:hints}, we have extracted a portion of the decompilation produced in \autoref{fig:hexrays-stripped} where Hex-Rays was not provided with correct types or names. In this case \texttt{a1} has been incorrectly typed as \texttt{int *} rather than \texttt{dict\_t *}. In \autoref{fig:types}, we show the ground truth types in this example. Notice that because \texttt{a1} is mistyped, accesses to fields in the \texttt{dict\_t} are represented with raw pointer arithmetic instead of field access notation. Although the code in \autoref{fig:hints} looks strange, each labeled hint reveals information about the true type of \texttt{a1}:

\begin{enumerate}
    \itemsep0em
    \item We can observe that the first two fields are integers that are directly compared and likely related to each other.
    \item The first comparison leads to an early return, suggesting some sort of bounds check.
    \item The first field is used as an index to another field in \texttt{a1} for assignment.
    \item There are in fact \textit{two} fields where \texttt{a1} is used as an index.
    \item The first field gets incremented during this function.
\end{enumerate}

Put together, an experienced reverse engineer could deduce that \texttt{a1} is a structure where the first two fields represent integer indexes, the latter being a maximum or capacity. Additionally, the structure contains two more fields which point to arrays, both of the same size which are accessed together. These \textit{hints}, when aggregated, could lead the reverse engineer to accurately reconstruct the type and names of these variables.

\smallsec{Have We Seen This Before?}
While one could implement handcrafted rules or heuristics to detect some of these patterns---as indeed many decompilers do---we instead turn to statistical techniques. Specifically, we observe that many variables have unique characteristics or \textit{usage signatures} that make them recognizable in other situations.

Our goal is therefore to find a way to represent these \textit{usage signatures} such that we can identify other variables with similar signatures. Given a large enough training corpus, we could then make prospective predictions: for a variable we want to predict the type or name of, find a variable in the training corpus that has the most similar signature. If two variables have very similar signatures, they probably have the same name and/or type.

\smallsec{N-Grams Generalize Usage Signatures.}
We implement this notion of \textit{usage signatures} with N-grams. In particular, we store contiguous sequences of tokens immediately preceding and following each location where a variable is mentioned in the decompilation. We can then search for other variables (in the training data) that occur within the same token sequences.

\section{{\sf \small STRIDE}'s Architecture}

Using the intuition discussed in Section~\ref{sec:rev}, we arrived at \sysname, a simple model for variable retyping and renaming. In this section we formalize concepts discussed previously and explain the architecture of this system in more detail. In subsequent sections we will evaluate the performance against prior work and discuss the implications.

\subsection{Decompiler Tokens}

For each function in a binary, a decompiler will produce a sequence of \textit{tokens}. Concatenated together the tokens resemble C source code, however this code is not guaranteed to compile or be syntactically correct. In fact, many decompilers introduce custom notation to assist the reverse engineer such as explicit sign- or zero-extension.

As part of its internal analysis, the decompiler will predict the location and types of local variables, which may be located on the stack or held in registers. Modern decompilers, such as Hex-Rays, perform a small amount of static analysis to try to infer the size and type of the variables, often limited to categorizing between several numeric types, pointer types, and identifying arrays. Without debug information, decompilers will typically generate placeholder names such as \texttt{v1}, \texttt{v2}, \texttt{a3}, etc. Hex-Rays uses heuristics to name variables in certain situations; for example, it may generates the name \texttt{result} for a variable that is returned from a function.

\subsection{N-Grams}

The concept of an \textit{N-gram} dates back at least to the 1940's~\cite{shannon1948mathematical}. Put simply, an \textit{N-gram} is a tuple of \textit{N} elements---in our case, these are \textit{decompiler tokens}. N-grams have been widely applied in the field of natural language processing as a way to perform fuzzy matching of text among other tasks. 

In our system, we use N-grams to model the \textit{usage signature} of a variable. Specifically, for a given variable in a sequence of decompiler tokens, we only consider the N-grams that immediately precede or follow the variable token.

\begin{figure}
    \resizebox{\columnwidth}{!}{
        \includegraphics{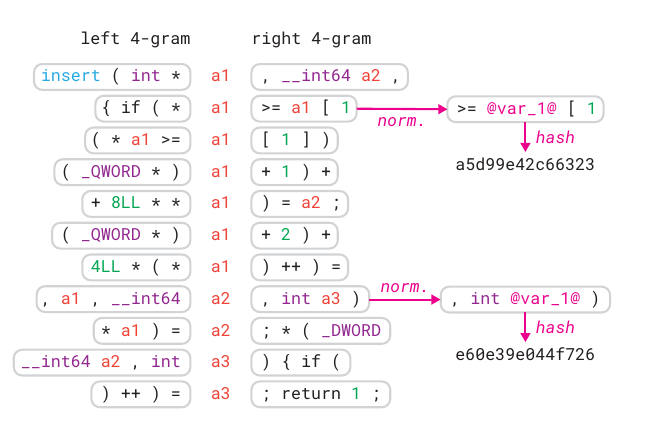}
    }
    \caption{All of the 4-grams for variables found in \autoref{fig:hexrays-stripped} and two examples of N-gram normalization and hashing.}
    \label{fig:ngram-hash}
\end{figure}

For example, \autoref{fig:ngram-hash} shows a list of all of the left and right \textit{4-grams} for variables found in the function listed in \autoref{fig:hexrays-stripped}. Every time a variable appears in the decompilation (i.e. the token \texttt{a1}, \texttt{a2}, or \texttt{a3}), we capture \textit{N} tokens (in this case 4) on either side as an N-gram.

During inference, we try to find the \textit{largest} N-grams surrounding a variable that exist in our training corpus. The intuition is that larger N-grams represent a more precise context, and therefore a more likely true match. Separately looking to the left and right of the variable allows for finding a precise match on one side but not the other. In some cases (particularly in optimized code) we find that variables may be found with common usage patterns to the left, but very different patterns to the right (or vice-versa).

\subsection{Normalization}
\label{sec:normalization}

Matching two N-grams requires the tokens to be identical. If the decompilation contains a lot of unique tokens, it will cause N-grams to be different in different contexts, even if the variables are actually the same. Therefore we apply a small amount of normalization before comparing N-grams.

\subsubsection{DIRT and DIRE} Two of the datasets we evaluate on, \work{DIRT}~\cite{chen2022augmenting} and \work{DIRE}~\cite{lacomis2019dire}, already contain some pre-normalization of the decompiler tokens. Specifically, in these datasets all strings literals are replaced with the \texttt{String} token and all number literals are replaced with the \texttt{Number} token. As a result, N-grams containing numbers and strings will match other N-grams with the same order of tokens but different numerical or string values.

\subsubsection{Address Leakage} In the \work{VarCorpus}~\cite{varbert} dataset we also evaluate on, there are a lot of cases where binary addresses end up in the token stream. For example, there are tokens like \texttt{LAB\_001042e9} indicating a label at address \texttt{1042e9}. These addresses can also appear as raw number literals (i.e. \texttt{0x114b28}), particularly in the splits that use the Ghidra decompiler. This address leakage results in many unique tokens which can prevent N-grams from matching. In order to preserve matching, we identified 25 token prefixes (such as \texttt{LAB\_}, \texttt{FUN\_}, \texttt{DAT\_}) that are followed by addresses and convert these tokens to a single representative token for each prefix. Additionally, we convert number literals above \texttt{0x100} into one of several tokens depending on the magnitude of the number. For example, \texttt{0x1234} becomes \texttt{NUM\_4}, \texttt{0x114b28} becomes \texttt{NUM\_6}, and so on. This conversion allows N-grams to match the magnitude of numbers without needing to precisely match literals like addresses.

\subsubsection{Canonicalization} Finally, the variable names captured in N-grams are an artifact of where the code was found in the function and which placeholder name the decompiler chose to use. We don't want to distinguish between say \texttt{a1 * a2} and \texttt{z4 * z5}. But we \textit{do} want to distinguish those from an N-gram like \texttt{r3 * r3} (the difference here is that it is the \textit{same} variable both times). Thus, we standardize the variable names in the N-gram. The first unique name becomes \texttt{@var\_1@}, the second unique name becomes \texttt{@var\_2@}, and so on. In the previous example, the first two N-grams would become \texttt{@var\_1@ * @var\_2@} while the third N-gram would become \texttt{@var\_1@ * @var\_1@}. Figure \autoref{fig:ngram-hash} shows two other examples of this variable name normalization.

\subsection{N-gram Database}
\label{sec:ngram-db}

Training the \sysname model requires building an N-gram database that maps N-grams found in training data to the most common variable names and types associated with those N-grams. This database is analogous to the weights in a conventional machine learning model and takes up a few gigabytes of memory.

In our implementation, we store the top 5 names and/or types for each N-gram along with the count of how many times that N-gram was associated with the particular name or type. Empirically, we find that it is useful to keep track of multiple labels for each N-gram, but beyond the top 5, one quickly hits diminishing returns.

In order to save space and enable efficient lookup, we don't store the full N-grams in this database, but rather a truncated hash. Specifically, for each N-gram, we construct a byte string by concatenating the tokens along with a separator character \texttt{\textbackslash xff} and appending the discriminator \texttt{left} or \texttt{right} (depending on the origin of the N-gram). We then compute the \texttt{SHA256} hash of this byte string and store the first 12 bytes in the database entry.

\subsection{Predictions}

\sysname compares the N-grams found in a prospective function to those found in the N-gram database in order to make predictions. The goal is to find variables in the database that have very similar N-grams and hence very similar \textit{usage signatures}. Specifically, we consider three factors to design our scoring system:

\begin{enumerate}
    \item The \textit{larger} the N-gram that matches, the more confident we can be about similarity.
    \item As \textit{more instances} of a variable have N-grams that match within a function, we gain more confidence that the match is correct.
    \item If we find an N-gram which occurs very frequently with only one name or type, we gain confidence that this is a very distinctive signature and hence, a strong match.
\end{enumerate}

\begin{figure*}[ht]
    \resizebox{\linewidth}{!}{
      \includegraphics{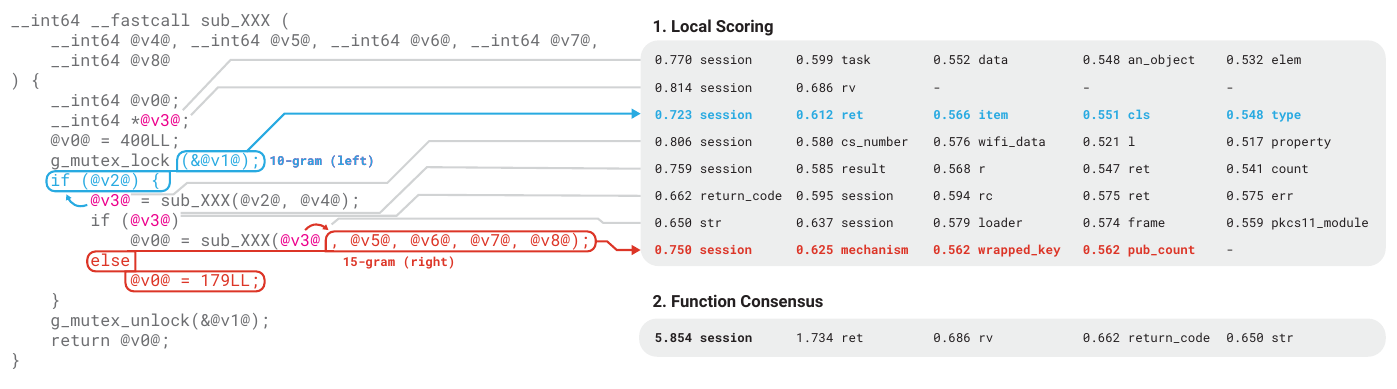}
    }
    \caption{An example variable name prediction with \sysname for the variable \texttt{v3}. First, \sysname finds the largest N-grams matching the areas next to instances of the variable and looks up the top 5 variable names. Next, \sysname sums the scores for each instance variable across the function. For this function, the top predicted name is \texttt{session} with a score of 5.854.}
    \label{fig:architecture}
\end{figure*}

We implement these factors directly with a two-part procedure as shown in \autoref{fig:architecture}.

\smallsec{Part 1: Local Scoring.}
For every location that a variable appears in a given function, we find the largest N-gram on the left and right of the variable that exist in the N-gram database. In our implementation, each N-gram may match up to five names or types. However, it is possible for N-grams to not match anything found in the database or to return fewer than five results (if fewer than 5 unique types or names were found in the training data with this N-gram).

For the labels returned, we assign a score based on the frequency that this label appears in the training data associated with this N-gram. For example, we may find an N-gram that maps to the following five variable names (with counts): (\texttt{this}, 360); (\texttt{data}, 148); (\texttt{new\_buffer}, 138); (\texttt{ctx}, 137); (\texttt{self}, 108).
For each label, we compute the percentage of samples in the top 5 that it occupies (\ie the specificity of the match) and map the result into the interval $[0.5, 1]$. In this case, we would assign the following local scores for these names:  1)~\texttt{this}: 0.702;
2)~\texttt{data}: 0.583; 3)~\texttt{new\_buffer}: 0.577; 4)~\texttt{ctx}: 0.577; 5)~\texttt{self}: 0.561.

Mapping the result into the interval $[0.5, 1]$ instead of using the raw ratio prevents discounting lower frequency types that appear in many places (i.e. factor 2). The intuition is that low frequency matches (a score of 0.5) in two different variable locations can ``overrule'' a high frequency match (score of 1) in a single variable location.

\smallsec{Part 2: Function Consensus.}
After computing local scores for each variable location, we aggregate the results and sum the scores for each label across all instances of a variable within the function. Ties are broken by label frequency in the training data. For example, if \texttt{buffer} and \texttt{buffer\_str} are tied as options for the variable name, \texttt{buffer} will be picked since it appears more times in the training data.

\subsection{Implementation}

Our implementation of \sysname is open source and consists of roughly 1,000 lines of Python. It is designed to work with the formats of the \work{DIRT}, \work{DIRE}, and \work{VarCorpus} datasets but can be easily adapted to other formats. In principle, any dataset with decompiler tokens and ground truth labels can be used.

Using \sysname requires first building the N-gram database as described in Section~\ref{sec:ngram-db}. Once constructed, the database can be used to make new predictions. By default, we use N-grams for $N \in \{60,30,15,14,13,12,11,10,9,8,7,6,5,4,3,2\}$ and we store the top 5 results for each N-gram when constructing the database. There is no particular magic to these sizes; however, in early tests, we found it was useful to have some large N-grams for highly precise matches and a wide range of smaller N-grams for less precise matches.

\section{Evaluation}

Our evaluation strategy is to compare (as fairly as possible) \sysname to the top-performing systems from prior work on both tasks, variable renaming and variable retyping.
We evaluate \sysname on the \work{DIRT}~\cite{chen2022augmenting} dataset for both variable retyping and renaming, to compare against the prior work \dirty system which can be applied to both tasks. Additionally, we evaluate \sysname on the DIRE~\cite{lacomis2019dire} and \work{VarCorpus}~\cite{varbert} datasets for variable renaming, to compare against the prior work systems \work{DIRE} and state-of-the-art \varbert.

\subsection{Experimental Setup}

We train and evaluate a separate version of \sysname for each dataset (and for each split of \work{VarCorpus}). For results that can be directly compared with prior work, we copy the results from respective publications. For each dataset, \sysname is provided \textit{only} the data in the training set. \sysname is not provided with data in the validation set (for \work{DIRE}/\work{DIRT}) or any pre-training data (for \work{VarCorpus}).

For comparisons with \dirty, we use the pre-trained model provided by the original authors in their GitHub repository.
\footnote{\href{https://github.com/CMUSTRUDEL/DIRTY}{https://github.com/CMUSTRUDEL/DIRTY}} 
\dirty can use beam search to improve accuracy at the cost of longer runtime. We evaluated \dirty with both $beam=0$, which is the default configuration in the GitHub repository and suggested configuration in the README, and at $beam=5$, which is the configuration used in the original publication~\cite{chen2022augmenting}. Note that when the beam size is not explicitly stated, we use $\text{DIRTY}_{beam=5}$ to provide a more fair comparison.

For comparisons with \varbert, we use the fine-tuned models published by its original authors in their GitHub repository for each dataset split.
\footnote{\href{https://github.com/sefcom/VarBERT}{https://github.com/sefcom/VarBERT}}
Predictions with \varbert and \dirty were made on a server with a NVIDIA L4 GPU.

\subsection{Datasets}

\begin{table}
  \footnotesize
  \centering
  \caption{Statistics of the \work{DIRT}, \work{DIRE}, and \work{VarCorpus} datasets. \work{VarCorpus} entries are summed across the 16 dataset splits. $^*$\textit{Binaries in \work{VarCorpus} are executables or libraries, not object files. Generally these consist of many object files.}}
  \label{fig:datasets}
  \begin{tabularx}{\columnwidth}{Xrrr}
    \toprule
    Feature & \work{DIRT}~\cite{chen2022augmenting} & \work{DIRE}~\cite{lacomis2019dire} & \work{VarCorpus}~\cite{varbert} \\
    \midrule
    \textbf{Binaries} \\
    \texttt{TRAIN} & 75,656 & 41,629 & 75,892$^*$ \\
    \texttt{VAL} & 9,457 & 8,048 & - \\
    \texttt{TEST} & 9,457 & 8,152 & 26,688$^*$ \\
    \midrule
    \textbf{Functions} \\
    \texttt{TRAIN} & 738,158 & 317,638 & 14,146,552 \\
    \texttt{VAL} & 142,048 & 64,447 & - \\
    \texttt{TEST} & 142,193 & 69,232 & 3,545,159 \\
    \midrule
    \textbf{Task} \\
    Renaming & yes & yes & yes \\
    Retyping & yes & - & - \\    
    \bottomrule
  \end{tabularx}
\end{table}

We evaluate \sysname on three datasets: \work{DIRE}, \work{DIRT}, and \work{VarCorpus}. \autoref{fig:datasets} provides statistics on the sizes of these datasets. In the rest of this section, we provide some more detail about each dataset.

\smallsec{DIRT.}
This dataset was introduced with \dirty~\cite{chen2022augmenting} as a benchmark for variable retyping and renaming. It consists of 94,570 binaries sourced from 8,878 randomly selected GitHub repositories (almost all C with a small fraction of C++). These binaries were randomly distributed into \texttt{TRAIN}/\texttt{VAL}/\texttt{TEST} sets with the ratio 72/14/14. Binaries were compiled with a custom tool, GHCC (Github Cloner and Compiler),
\footnote{\href{https://github.com/CMUSTRUDEL/ghcc}{https://github.com/CMUSTRUDEL/ghcc}}
which used GCC 9.2.0 and no optimizations (\texttt{-O0}) internally.

Binaries in this dataset were compiled with DWARF debug information. Hex-Rays then decompiled both the binary with debug information and a \textit{stripped} version of the binary without debug information. The resulting variables in each of the functions were compared to generate an alignment, providing a ground-truth type label for variables.

\smallsec{DIRE.}
This dataset is a slightly older dataset developed for \work{DIRE}~\cite{lacomis2019dire} to evaluate variable renaming accuracy. It consists of fewer total binaries---only 57,829---but was prepared in a similar manner to the more recent \work{DIRT} dataset.

The \work{DIRE} model incorporates both lexical and structural information extracted from Hex-Rays to help inform variable names. As such, this dataset includes extra AST information in addition to the raw decompiler tokens. However, for integration with \sysname we ignore this extra information and use \textit{only} the decompiler tokens.

\smallsec{VarCorpus.}
The \work{VarCorpus} dataset is the newest and largest dataset for variable renaming, developed for \work{VarBERT}~\cite{varbert}. The dataset consists of C and C++ packages sourced from the Gentoo repository. Packages were compiled at four different optimization levels (\texttt{-O0}, \texttt{-O1}, \texttt{-O2}, \texttt{-O3}) and two different decompilers were used to generate decompilation: IDA Hex-Rays and Ghidra. The resulting entries were then split into a \texttt{TRAIN} and \texttt{TEST} set either by \textit{function}, where each function is randomly assigned to \texttt{TRAIN} or \texttt{TEST}, or by \textit{binary} where all functions from the same binary are either all in the \texttt{TRAIN} set or all in the \texttt{TEST} set. In total there are 16 combinations of decompiler/optimization/split.

\subsection{Measuring Accuracy}

Similar to prior work, we treat the variable renaming and retyping tasks as a classification problem. For each variable, the objective is to predict the correct name or type \textit{exactly}. This means that in practice a name prediction of \texttt{size} where the ground truth name is \texttt{len} would be marked incorrect even though the meaning is similar. While this metric may underestimate the performance of these models in practice, it provides an unambiguous way to compare performance in automated evaluations like ours.

There is also a slight difference in how accuracy is computed between datasets (following prior work). On the \work{DIRT} and \work{DIRE} datasets, a prediction is made for every unique variable in a function and it is counted as correct if it matches the ground truth label. Specifically, if a function has a variable \texttt{a1} which appears in 5 places, only one prediction is made for this variable and it is counted as one result. The total accuracy for the dataset is the average accuracy across all variables.

On the \work{VarCorpus} dataset, a prediction is made for every \textit{instance} of a variable in a function. If a function has a variable \texttt{a1} appearing in 5 places, there are 5 predictions made for this variable (one at each location) and each instance is counted as one result. The total accuracy for the dataset is the average accuracy across all variable \textit{instances}. With \sysname, we still make a single prediction per variable on the \work{VarCorpus} dataset but we weight the prediction accuracy based on variable frequency so as to compare fairly.

\section{Results}

In this section, we report results from various experiments to evaluate \sysname on the \work{DIRT}, \work{DIRE}, and \work{VarCorpus} datasets.

\smallsec{N-gram Size.}
We performed an initial experiment on the \work{DIRT} dataset to see how frequently N-grams match between the \texttt{TRAIN} and \texttt{TEST} set and how accurate the match is. Specifically, for each variable in the \texttt{TEST} set, we retrieve N-grams of various sizes and check if the N-gram matches one we've seen in the \texttt{TRAIN} set (\% Matched). If it does, we then check if the corresponding entry in the database contains the target variable name in the Top-1, Top-3, or Top-5 entries (i.e. the accuracy of this match).

We report these results in \autoref{fig:nsize}. Small N-grams are almost always found in the database (high \% matched), but the resulting match is not very accurate. On the other hand, large N-grams are less likely to match something we've seen before, but if they do, it is much more likely to be a correct match.

\begin{table}
  \footnotesize
  \centering
  \caption{N-gram matching frequency and accuracy for the \work{DIRT} dataset.}
  \label{fig:nsize}
  \begin{tabular}{rcccc}
    \toprule
    & & \multicolumn{3}{c}{Accuracy (\%)} \\
    \cmidrule{3-5}
    N & Matched (\%) & Top-1 & Top-3 & Top-5 \\
    \midrule
    60 & 70.8 & 64.2 & 66.6 & 67.1 \\
    30 & 74.6 & 63.6 & 67.6 & 68.5 \\
    15 & 83.6 & 54.0 & 62.9 & 65.5 \\
    14 & 84.9 & 52.2 & 61.7 & 64.5 \\
    13 & 86.2 & 50.1 & 60.1 & 63.3 \\
    12 & 87.6 & 47.7 & 58.2 & 61.7 \\
    11 & 89.1 & 45.0 & 55.9 & 59.8 \\
    10 & 90.6 & 41.9 & 53.3 & 57.4 \\
    9 & 92.0 & 38.5 & 50.0 & 54.4 \\
    8 & 93.4 & 34.8 & 46.1 & 50.8 \\
    7 & 94.6 & 30.8 & 41.8 & 46.5 \\
    6 & 96.0 & 27.1 & 37.4 & 42.1 \\
    5 & 96.9 & 23.3 & 32.9 & 37.5 \\
    4 & 97.7 & 19.6 & 28.5 & 32.9 \\
    3 & 98.5 & 16.2 & 24.2 & 28.1 \\
    2 & 99.3 & 13.4 & 20.6 & 24.4 \\
    \bottomrule
  \end{tabular}
\end{table}

\smallsec{Variable Retyping.}
\autoref{fig:acc-retyping} compares the accuracy of \sysname and \dirty on variable retyping. Chen et al.\ \cite{chen2022augmenting} split the dataset three-fold:
\begin{itemize}
  \item \textbf{In Train}: Functions that appear in their entirety in the \texttt{TRAIN} set. For example, this set is primarily shared library code that is linked statically.
  \item \textbf{Not In Train}: Functions not appearing in the \texttt{TRAIN} set.
  \item \textbf{Overall}: All functions, without distinguishing.
\end{itemize}

In the DIRT \texttt{TEST} set, roughly 60\% of functions are \textit{In Train} and 40\% of functions are \textit{Not In Train}.

Unsurprisingly, \sysname achieves close to 100\% accuracy on the \textit{In Train} subset, as it can use large N-grams to memorize entire subsections of functions. Interestingly, \sysname also achieves nearly 14.1\% more accuracy on \textit{Not In Train} functions than \dirty. The 1.5\% accuracy loss on the \textit{In Train} set is caused by the occasional small function that is token-for-token identical to another function except for target types.

\begin{table}
  \caption{\sysname vs \dirty on the retyping task.}
  \label{fig:acc-retyping}
  \footnotesize
  \begin{tabularx}{\columnwidth}{Xlccc}
    \toprule
    Dataset & Model & Overall & In Train & Not In Train \\
    \midrule
    \multirow{3}{*}{\work{DIRT}} & $\text{\dirty}_{beam=0}$~\cite{chen2022augmenting} & 73.9 & 87.5 & 51.4 \\
    & $\text{\dirty}_{beam=5}$~\cite{chen2022augmenting} & 74.8 & 88.6 & 52.3 \\
    & \sysname & \textbf{85.0} & \textbf{98.5} & \textbf{66.4} \\
    \bottomrule
  \end{tabularx}
\end{table}

\smallsec{Variable Renaming.}
\autoref{fig:acc-rename} compares the variable renaming accuracy of \sysname to other state-of-the-art systems on the \work{DIRT} and \work{DIRE} datasets. We report accuracy for the same categories as described in the previous section.

Similar to variable retyping, \sysname is quite effective at variable renaming. It achieves a near perfect accuracy on the \textit{In Train} subset on both datasets, and on the \textit{Not In Train} subsets it comes very close to matching \varbert  of the older \work{DIRE} dataset and outperforms \varbert considerably on the larger (and newer) \work{DIRT} dataset. 

One can argue that evaluating on the \textit{Not In Train} subsets for \work{DIRE} and \work{DIRT} is more important, as a high performance on the \textit{In Train} subset can be achieved simply by memorizing examples. In real-world applications, one is less likely to encounter identical functions to those seen in training data and so the \textit{Not In Train} subset is more representative. For completeness, we report results for all splits in order to more directly compare with prior work.

Overall, we find that \sysname is both capable of memorizing examples, as demonstrated by high accuracy on the \textit{In Train} subset, and generalizing with SOTA performance, as demonstrated by competitive accuracy on the \textit{Not In Train} subset.

\autoref{fig:acc-varcorpus} compares the variable renaming accuracy of \sysname and \varbert on the \work{VarCorpus} dataset across all 16 splits. \sysname achieves higher accuracy on 11 of the splits with an average accuracy improvement of 1.55\%. Interestingly, \sysname considerably outperforms \varbert on \texttt{-O0} splits for both IDA and Ghidra with an average accuracy improvement of 5.4\%. Likely, the regularities introduced by \texttt{-O0} generate the same token sequences more consistently compared to higher optimization levels which may optimize away parts of code.

\begin{table}
  \footnotesize
  \caption{Accuracy on the renaming task. Non-\sysname data is reproduced from the respective publications.}
  \label{fig:acc-rename}
  \begin{tabularx}{\columnwidth}{Xlccc}
    \toprule
    Dataset & Model & Overall & In Train & Not In Train \\
    \midrule
    \multirow{4}{*}{\work{DIRT}} & \work{DIRE}~\cite{lacomis2019dire} & 57.5 & 75.6 & 31.8 \\
    & $\text{\dirty}_{beam=5}$~\cite{chen2022augmenting} & 66.4 & 87.1 & 36.9 \\
    & \work{VarBERT}~\cite{varbert} & - & - & 51.3 \\
    & \sysname & \textbf{80.9} & \textbf{98.8} & \textbf{56.2} \\
    \midrule
    \multirow{6}{*}{\work{DIRE}} & \work{DIRE}~\cite{lacomis2019dire} & 74.3 & 85.5 & 35.3 \\
    & $\text{\dirty}_{beam=5}$~\cite{chen2022augmenting} & 81.4 & 92.6 & 42.8 \\
    & \work{DIRECT}~\cite{direct} & - & - & 42.8 \\
    & \work{HexT5}~\cite{hext5} & - & 90.0 & 55.0 \\
    & \work{VarBERT}~\cite{varbert} & - & - & \textbf{61.5} \\
    & \sysname & \textbf{90.1} & \textbf{99.8} & 61.1 \\
    \bottomrule
  \end{tabularx}
\end{table}

\begin{table}
    \footnotesize
    \caption{Accuracy of \work{VarBERT} and \sysname on the renaming task evaluated on the \work{VarCorpus} dataset.}
    \label{fig:acc-varcorpus}
    \begin{tabularx}{\columnwidth}{Xcccc|c}
      \toprule
        Decompiler & Opt & Split & \work{VarBERT}~\cite{varbert} & \sysname & $\Delta$ \\
      \midrule
        \multirow{8}{*}{IDA} & \multirow{2}{*}{O0} & function & 54.01 & \textbf{63.37} & +9.36 \\
        & & binary & 44.80 & \textbf{49.08} & +4.28 \\
        & \multirow{2}{*}{O1} & function & 53.51 & \textbf{55.33} & +1.82 \\
        & & binary & \textbf{42.55} & 42.42 & -0.13 \\
        & \multirow{2}{*}{O2} & function & 54.43 & \textbf{55.55} & +1.12 \\
        & & binary & 42.40 & \textbf{43.46} & +1.06 \\
        & \multirow{2}{*}{O3} & function & 56.00 & \textbf{56.86} & +0.86 \\
        & & binary & 40.70 & \textbf{42.30} & +1.60 \\

      \midrule

        \multirow{8}{*}{Ghidra} & \multirow{2}{*}{O0} & function & 60.13 & \textbf{64.40} & +4.27 \\
        & & binary & 46.10 & \textbf{49.80} & +3.70 \\
        & \multirow{2}{*}{O1} & function & \textbf{58.47} & 56.33 & -2.14 \\
        & & binary & \textbf{42.87} & 41.22 & -1.65 \\
        & \multirow{2}{*}{O2} & function & \textbf{54.49} & 54.44 & -0.05 \\
        & & binary & 40.49 & \textbf{41.46} & +0.97 \\
        & \multirow{2}{*}{O3} & function & \textbf{54.68} & 54.33 & -0.35 \\
        & & binary & 40.09 & \textbf{40.20} & +0.11 \\
      \bottomrule
    \end{tabularx}
  \end{table}

\subsection{Resource Requirements}

In \autoref{fig:resources} we compare the prediction speed of \sysname, \dirty, and \varbert. Both \dirty and \varbert are transformer-based models that are designed to run on a GPU. \sysname however, runs entirely on the CPU as the main inference step is a simple database lookup rather than a feed-forward pass of a neural model. We report the prediction speed (ms per function) on both a NVIDIA L4 GPU (for the transformer models) and on the CPU for a direct comparison.

Notably, \sysname is considerably faster than both models even without the use of GPU: more than 5 times faster than \varbert and more than 25 times faster than \dirty when beam search is not used.

\begin{table}
  \footnotesize
  \caption{Prediction speed comparison.} %
  \label{fig:resources}
  \begin{tabularx}{\columnwidth}{XXrr}
    \toprule
    Mode & Method & Time/func. (ms) \\
    \midrule
    \multirow{3}{*}{GPU (NVIDIA L4)} & $\text{\dirty}_{beam=0}$ & 208.3 \\
    & $\text{\dirty}_{beam=5}$ & 1134.5 \\
    & \varbert & 43.7 \\
    \midrule
    \multirow{4}{*}{CPU} & $\text{\dirty}_{beam=0}$ & 1723.7 \\
    & $\text{\dirty}_{beam=5}$ & 8476.3 \\
    & \varbert & 351.1 \\
    & \sysname & \textbf{8.2} \\
    \bottomrule
  \end{tabularx}
\end{table}

\smallsec{Function Size.}
A current limitation of transformer models is the input size. \dirty uses a fixed input size of 512 tokens and functions larger than this are truncated. For large functions, this means \dirty only has partial information about the tokens in a given function. Similarly, \varbert has a max input size of 800 tokens and splits functions into multiple parts to make predictions separately.

\begin{figure*}[t]
\centering
  \resizebox{0.8\textwidth}{!}{
    \includeinkscape{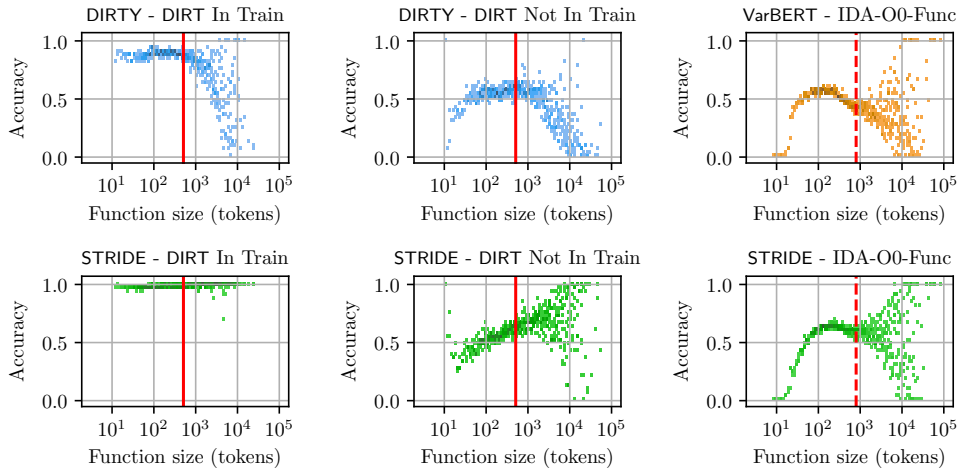}
  }
  \caption{Effect of function size on variable renaming accuracy. Left column: \sysname (bottom) compared to \dirty (top) on the \work{DIRT} In Train set. Middle column: \sysname (bottom) compared to \dirty (top) on the \work{DIRT} Not In Train set. Right column: \sysname (bottom) compared to \varbert (top) on the \work{VarCorpus} IDA-O0-Func split. Solid red line: \dirty max input (512 tokens), dashed red line: \varbert max input (800 tokens). In each chart, functions were aggregated into 200 evenly spaced logarithmic buckets from $10^{0.5}$ to $10^5$ and we display a 2D histogram of those buckets.}
  \label{fig:function-size}
\end{figure*}

In \autoref{fig:function-size} we show how function size affects prediction accuracy. \dirty exhibits a significant drop in performance after hitting the transformer input window (solid red line). \sysname does not suffer from this problem since it does not truncate functions and virtually ``attends'' to all variable usage sites (left and middle column). There is a similar albeit less strong effect for \varbert (right column) after hitting its max input size (dashed red line).

\smallsec{Name Frequency.}
In \autoref{fig:type-frequency}, we evaluate the renaming accuracy of \sysname, \dirty, and \varbert based on how frequently the \textit{target name} appears in the training data. Specifically, we compare \sysname and \dirty on the \work{DIRT} dataset (both In-Train and Not-In-Train splits) and \sysname and \varbert on the IDA/\texttt{-O0}/function split of the \work{VarCorpus} dataset.

Unsurprisingly, all models are better at predicting names that have been seen more frequently. \sysname outperforms both \dirty and \varbert for all frequency ranges and this effect is especially notable for names that appear fewer than 100 times in the training data. On the \work{DIRT} Not-In-Train split, \sysname has 32\% higher accuracy on types seen 1-10 times and on the \work{VarCorpus} IDA-O0-Func split, \sysname achieves more than 35\% accuracy on the same frequency range while \varbert does not predict any correctly.

\begin{table}
  \footnotesize
  \caption{Renaming accuracy of \sysname (S), \dirty (D), and \varbert (V) based on frequency of the name in training data.}
  \label{fig:type-frequency}
  \begin{tabularx}{\columnwidth}{Xcccccc}
    \toprule
    & \multicolumn{4}{c}{\work{DIRT}} & \multicolumn{2}{c}{\work{VarCorpus}} \\
    \cmidrule(lr){2-5} \cmidrule{6-7}
    & \multicolumn{2}{c}{In Train} & \multicolumn{2}{c}{Not In Train} & \multicolumn{2}{c}{IDA-O0-func} \\
    \cmidrule(lr){2-3} \cmidrule(lr){4-5} \cmidrule{6-7}
    Name Frequency & D & S & D & S & V & S \\
    \midrule
    1 & 0.0 & \textbf{100.0} & 0.0 & \textbf{21.1} & 0.0 & \textbf{12.8} \\
    1-10 & 28.0 & \textbf{98.8} & 8.5 & \textbf{40.5} & 0.0 & \textbf{35.4} \\
    11-100 & 68.8 & \textbf{99.1} & 29.6 & \textbf{56.1} & 18.8 & \textbf{53.8} \\
    101-1,000 & 84.2 & \textbf{98.8} & 43.8 & \textbf{55.1} & 51.9 & \textbf{60.8} \\
    1,001-10,000 & 88.7 & \textbf{98.8} & 52.0 & \textbf{53.2} & 58.8 & \textbf{61.9} \\
    10,001-100,000 & 86.5 & \textbf{98.9} & 55.2 & \textbf{53.6} & 63.8 & \textbf{64.9} \\
    100,001+ & 89.3 & \textbf{98.3} & 63.3 & \textbf{59.5} & 65.7 & \textbf{69.1} \\
    \bottomrule
  \end{tabularx}
\end{table}

\smallsec{Identifiers.}
In the \work{DIRT} dataset, while debug information was removed, identifiers such as function names and global variables were retained. These identifiers are often highly suggestive of variable types with names such as \texttt{av\_log} or \texttt{pci\_default\_write\_config}. In a real-world setting, identifiers may not be present.

To assess the usefulness of \sysname in such a setting, we trained a variant of \sysname where all identifiers were removed from the dataset and replaced with the single unknown token \texttt{?}. Specifically, we constructed a list of 67 built-in decompiler tokens such as \texttt{while}, \texttt{if}, \texttt{\{}, \texttt{;}, \texttt{void}, \texttt{+=}, etc... and replaced every token not in this list with \texttt{?}.

We report the accuracy of this stripped model $\text{\sysname}_{\text{strip}}$ compared to the original model in \autoref{fig:acc-identifiers}. As expected, the stripped version performs slightly worse than the original \sysname model but only by a few percentage points.

\begin{table}
  \footnotesize
  \caption{Accuracy of \sysname at variable renaming with and without identifiers.}
  \label{fig:acc-identifiers}
  \begin{tabularx}{\columnwidth}{Xccc}
    \toprule
    Predicting Names (\work{DIRT}) & Overall & In Train & Not In Train \\
    \midrule
    \sysname & 80.9 & 98.8 & 56.2 \\
    $\text{\sysname}_{\text{strip}}$ & 77.7 & 95.3 & 53.5 \\
    \bottomrule
  \end{tabularx}
\end{table}

\section{Discussion}

In this section, we discuss our results more broadly, reflecting on lingering limitations of all techniques and potential future directions for variable retyping and renaming.

\subsection{Looking Back}

Our proposed system \sysname is a less-is-more technique with demonstrated benefits, across several dimensions, when compared to the existing state-of-the-art transformer-based models \dirty and \varbert. \sysname matches or exceeds the accuracy of both in almost all scenarios on the variable renaming task, including when dealing with compiler optimizations, and far exceeds \dirty's accuracy on the variable retyping task, while being conceptually simpler (non-neural) and more interpretable, arguably easier to train, and close to an order of magnitude faster at prediction time on CPU compared to the fastest GPU-accelerated competitor \varbert. 

Reflecting on fundamental differences between our N-grams-based approach in \sysname and the two transformers \dirty and \varbert, one can first note that the input size of both transformers is quite limited (512 input tokens for \dirty and 800 for \varbert; \dirty only embeds the first 512 tokens of a function and truncates the remaining parts, if longer; \varbert splits functions into 800-token chunks and predicts separately all the instances of variables in them).  In addition, \dirty recursively queries a multi-task decoding network to predict variables and limits this step to 32 variables, \ie extra variables in a function are ignored.
In contrast, \sysname uses the tokens around the variable of interest to make local predictions which are then aggregated into a function-wide prediction. This process allows \sysname to scale to larger functions than \dirty and \varbert can embed.
In fact, as shown in \autoref{fig:function-size}, \sysname tends to perform better as the function gets larger, likely because there is a higher chance of finding a distinctive usage pattern for a variable. Additionally, variables in a given function are predicted independently and it is not necessary to artificially limit how many variables can be predicted at a time.

Compared to \dirty and \varbert, \sysname is probably also less affected by class imbalance. When training a classifier, if some classes appear more frequently than others, those classes tend to be more important to learn within the limited weight-space of the model. In practice, this means that without correction, ML models tend to be better at predicting high-accuracy classes at the expense of poor accuracy on low-frequency classes. However, as others have observed, N-gram based models do not have such a strong weakness, generally outperforming neural language models on low frequency words \cite{neubig2016generalizing}. Intuitively, even a type seen just once in the training data can have a very distinctive N-gram which can be recognized at inference time.
Our experiments with \dirty and \varbert, both quite small transformers by today's standards, suggest that could be happening here as well.

However, there are also fundamental limitations of \sysname. Notably, \sysname cannot fuzzy match for either task by construction and may be overly specific to certain token sequences, whereas transformer-based models like \dirty and \varbert could have learned to fuzzy match in theory (but don't seem to benefit much from this capability in our experiments, given \sysname's overall accuracy numbers). Still, the transformers could be further enhanced to learn more sophisticated semantic representations of variables, which are starting to emerge~\cite{wainakh2021idbench, chen2022varclr}, whereas \sysname cannot in the same ways.
It's also noteworthy that \sysname can't generate names / types out of vocabulary (\ie we can only predict names / types we've seen in training), whereas \dirty can in theory---although as others have noted, not to great effect~\cite{varbert}.

\subsection{Looking Forward}

 The field is already seeing many rapid advancements specifically with large language models (LLMs). Advancements to these architectures will surely be adapted to the field of decompilation. Indeed, in 2022, researchers explored the use of LLMs for reverse engineering related tasks and found that despite promising results, LLMs were not yet capable of zero-shot reverse engineering~\cite{pearce2022pop}.  One year later, multiple research groups have shown that advances in both LLMs~\cite{wu2023exploring} and prompting techniques~\cite{hudegpt} have significantly advanced the abilities of LLMs to decompile binary code and perform other software security tasks.

Certainly, an LLM-like hypothetical future transformer with larger attention window could outperform \sysname. In fact, we expect it will, and probably sooner than we think. \sysname is not a reason to stop training transformers altogether but rather a demonstration of the value of thinking deeply about the constraints imposed by the task to design systems that are simpler, cheaper, faster, etc. There are many scenarios where systems like \sysname might be preferable, even when a better-performing LLM becomes available for the same task. Some, like relatively fast runtime inference compared to an LLM, are obvious. Others are more subtle. For example, certain tasks in reverse engineering require more confidentiality, particularly in vulnerability research or malware reverse engineering. Potential users may be wary of uploading their data to remote servers to be processed by black-box models. Systems which can run quickly on standard commodity hardware may still be desired for these applications. There has been recent work in fine-tuning smaller models on the output of large models to run on consumer hardware~\cite{llama,alpaca,gpt4all}, however the most powerful LLMs are still only available as a service.

Overall, our work demonstrates that certain innate shortcomings of transformers can be circumvented by careful design and that perhaps the way forward is some fusion architecture combining the positive aspects of both models.

\section{Conclusion}

We introduced \sysname, a new statistical approach to variable retyping and renaming in decompiled code. Our approach is much simpler than state-of-the-art models while also being faster and more efficient. We discussed the design choices that lead to this improvement and demonstrated how limitations of a transformer-based architecture negatively affect performance on these tasks.

While variable retyping and renaming are not solved problems, \sysname takes a long, decisive step forward in the domain and demonstrates how simple techniques with the right design choices can outperform complex machine learning models. We have open-sourced our implementation and intend to pursue integration with modern decompilers for use in real-world reverse engineering tasks.

\section*{Availability}

Our implementation of \sysname is open-source and available at \url{https://github.com/hgarrereyn/STRIDE}.

\bibliographystyle{IEEEtran}
\bibliography{paper}

\end{document}